\documentclass{article}

\usepackage[utf8]{inputenc}
\usepackage[T1]{fontenc}
\usepackage{arxiv}
\usepackage{amsmath,amsfonts,amssymb}
\usepackage{graphicx}
\usepackage{float}
\usepackage{booktabs}
\usepackage[numbers]{natbib}
\usepackage[colorlinks=true, allcolors=blue]{hyperref}
\usepackage{authblk}
\usepackage{siunitx}
\usepackage{tikz}
\usetikzlibrary{arrows.meta}

\title{Resolution--Noise Characteristics of Common FDK Filter Kernels: A Practical Reference for Preclinical Cone-Beam Micro-CT}
\renewcommand{\shorttitle}{FDK Filter Kernel Reference for Micro-CBCT}

\author[a]{Falk L. Wiegmann}
\author[a,b]{Nancy L. Ford\thanks{Corresponding author: nlford@dentistry.ubc.ca}}
\affil[a]{Department of Physics and Astronomy, The University of British Columbia, Vancouver BC, Canada}
\affil[b]{Department of Oral Biological and Medical Sciences, The University of British Columbia, Vancouver BC, Canada}

\begin{document}

\maketitle

\begin{abstract}
The ramp filter kernel and cutoff frequency are fundamental parameters of the Feldkamp--Davis--Kress (FDK) algorithm that determine the resolution and noise characteristics of the reconstructed image. Despite their importance, systematic evaluations of their combined effect on task-based image quality in preclinical micro-CT are scarce, and many studies do not report the filter configuration used. We reconstruct identical data from a GE eXplore CT~120 scanner using four filter kernels (ramp, Shepp-Logan, cosine, Hamming) at four cutoff frequencies (1.0, 0.8, 0.6, and $0.379 \times$~Nyquist, matched to the detector-to-voxel size ratio) and evaluate each of the sixteen configurations using the modulation transfer function (MTF), noise power spectrum (NPS), and non-prewhitening detectability index (NPW~$d'$). Qualitative assessment is performed on a mouse lung specimen. Across the sixteen configurations, MTF$_{10}$ ranges from 0.93 to \SI{2.35}{lp/mm}, integrated NPS from 75\,670 to 13\,259~HU$^2$, and the Rose criterion crossing diameter from 2.86 to \SI{0.93}{mm} at $\Delta C = \SI{500}{HU}$ and from 7.74 to \SI{3.62}{mm} at \SI{100}{HU}. This note presents the data as a concise visual and quantitative reference for groups selecting FDK filter parameters for preclinical cone-beam CT.
\end{abstract}


%% ============================================================
\section{Introduction}\label{sec:introduction}
%% ============================================================

The Feldkamp--Davis--Kress (FDK) algorithm~\cite{feldkamp_fdk_1984} is the standard reconstruction method for cone-beam micro-CT~\cite{wiegmann_fdk_benchmark_2026}. A central step in the FDK pipeline is ramp filtering: each detector row is convolved with a filter whose ideal frequency response is $|f|$, compensating for the non-uniform sampling density inherent in the projection-slice geometry~\cite{ramachandran_ramp_1971, kak_slaney_1988}.

In practice, the ideal ramp is never used directly because its unbounded gain at high frequencies amplifies noise. Instead, the ramp is multiplied by an apodization window $W(f)$ that attenuates high-frequency content, and the result is truncated at a cutoff frequency $f_c$. Four classical windows are in common use: none (the pure ramp, also known as Ram-Lak~\cite{ramachandran_ramp_1971}), Shepp-Logan~\cite{shepp_logan_1974}, cosine, and Hamming~\cite{kak_slaney_1988}. Together, the window and cutoff determine the resolution and noise characteristics of the reconstructed image.

Despite the importance of these parameters, most preclinical micro-CT studies either do not report their filter configuration or select a single setting without systematic justification. This note provides a reference for filter selection by sweeping four windows across four cutoff frequencies on identical data and evaluating each configuration using the modulation transfer function (MTF), noise power spectrum (NPS), and non-prewhitening (NPW) detectability index $d'$~\cite{aapm_tg233_2019}. Filter and cutoff effects on cone-beam CT image quality have been examined previously~\cite{ghani_2017, choi_2020, han_baek_2018, houno_2017, lagerwerf_2020, tward_2008, gang_2014, bhattarai_2024}, but typically with a single image-quality metric or a single filter family in isolation. The novelty of this note lies only in combining MTF, NPS, and task-based detectability on identical scan data, in including the physically motivated matched cutoff $f_c = d_a / \Delta x$, and in presenting the full window--cutoff grid as a single visual reference.

%% ============================================================
\section{Background}\label{sec:background}
%% ============================================================

The FDK ramp filter in the frequency domain takes the form
\begin{equation}\label{eq:filter}
H(f) = |f| \cdot W(f), \quad |f| \leq f_c,
\end{equation}
where $f_c$ is the cutoff frequency and $W(f)$ is the apodization window. The four windows evaluated in this study are:
\begin{align}
\text{Ramp (Ram-Lak):} \quad & W(f) = 1, \label{eq:ramp} \\
\text{Shepp-Logan:} \quad & W(f) = \mathrm{sinc}\!\left(\frac{f}{2f_c}\right), \label{eq:shepp} \\
\text{Cosine:} \quad & W(f) = \cos\!\left(\frac{\pi f}{2f_c}\right), \label{eq:cosine} \\
\text{Hamming:} \quad & W(f) = 0.54 + 0.46\cos\!\left(\frac{\pi f}{f_c}\right). \label{eq:hamming}
\end{align}
These windows provide progressively stronger high-frequency attenuation, from no apodization (ramp) to substantial suppression (Hamming). All are zero for $|f| > f_c$.

The cutoff frequency $f_c$ is expressed as a fraction of the detector Nyquist frequency $f_\mathrm{det} = 1/(2 d_a)$, where $d_a$ is the detector pixel pitch. For the eXplore CT~120 scanner ($d_a = \SI{0.0284}{mm}$), $f_\mathrm{det} = \SI{17.6}{lp/mm}$. A physically motivated choice is the \emph{matched} cutoff $f_c = (d_a / \Delta x) \cdot f_\mathrm{det} = 1/(2\Delta x)$, where $\Delta x$ is the reconstruction voxel size. We adopt $\Delta x = \SI{0.075}{mm}$ as our reference voxel spacing, for which $d_a / \Delta x = 0.379$ and $f_c = \SI{6.67}{lp/mm}$---exactly the reconstruction Nyquist frequency, beyond which the voxel grid cannot represent additional detail.

%% ============================================================
\section{Methods}\label{sec:methods}
%% ============================================================

\subsection{Scanner and Acquisition}\label{sec:scanner}

All data were acquired on the GE eXplore CT~120 micro-CT scanner at \SI{80}{kVp}, \SI{40}{mA}, with \SI{16}{ms} exposure per frame and no frame averaging. The flat-panel detector comprises $3500 \times 2300$ pixels with a pixel pitch of \SI{0.0284}{mm}. Table~\ref{tab:geometry} summarises the scanner geometry. A short-scan acquisition (\ang{193}, 220 projections) was used for all reconstructions.

\begin{table}[H]
\centering
\caption{Scanner geometry parameters for the eXplore CT~120.}
\label{tab:geometry}
\small
\begin{tabular}{lc}
\toprule
Parameter & Value \\
\midrule
Detector size (pixels) & $3500 \times 2300$ \\
Detector pixel pitch (mm) & 0.0284 \\
Source-to-detector distance (mm) & 451.5 \\
Source-to-isocentre distance (mm) & 396.4 \\
Scan arc & \ang{193} ($180^{\circ} + 2\gamma_m$) \\
Number of projections & 220 \\
\bottomrule
\end{tabular}
\end{table}

\subsection{Specimens}\label{sec:specimens}

Quantitative image quality was evaluated on the mCTP~610 image quality phantom (Shelley Medical Imaging Technologies, Canada), which contains a slanted-edge insert (air/acrylic boundary) for MTF measurement and a homogeneous water region for NPS characterisation. Visual comparison was performed on a mouse lung specimen acquired with respiratory gating as part of a previously published study~\cite{ford_respiratory_2025} (see Ethics Statement).

\subsection{Reconstruction}\label{sec:reconstruction}

Reconstructions were performed using our open-source FDK pipeline~\cite{wiegmann_fdk_benchmark_2026} on a $1247 \times 1247 \times 933$ grid with isotropic \SI{0.075}{mm} voxels. The pipeline processes each projection $p(\beta, u, v)$ at gantry angle $\beta$ with detector coordinates $(u, v)$ through the following steps:
\begin{enumerate}
    \item \textbf{Pre-processing}: dark-current subtraction, flood-field normalisation, and logarithmic transform to convert raw detector intensities to line integrals of attenuation.
    \item \textbf{Cosine weighting}: multiplication by $D / \sqrt{D^2 + u^2 + v^2}$, where $D$ is the source-to-detector distance, to account for varying path lengths in cone-beam geometry.
    \item \textbf{Parker weighting}: for short-scan acquisitions, multiplication by smooth weighting functions that correct for non-uniform data redundancy near the scan-arc boundaries~\cite{parker_weights_1982, wiegmann_parker_2026}.
    \item \textbf{Ramp filtering}: one-dimensional convolution of each detector row with a filter kernel $H(f) = |f| \cdot W(f)$ as defined in Equation~\ref{eq:filter}, with window $W(f)$ and cutoff $f_c$ as described in Section~\ref{sec:background}.
    \item \textbf{Backprojection}: three-dimensional voxel-driven backprojection with distance weighting.
    \item \textbf{HU calibration}: conversion from linear attenuation to Hounsfield units using air and water reference values.
\end{enumerate}

\begin{figure}[H]
    \begin{center}
    \begin{tikzpicture}[
        box/.style={
            draw, rounded corners=3pt, minimum height=0.8cm,
            minimum width=2.0cm, align=center, font=\small,
            fill=blue!6, line width=0.6pt
        },
        highlightbox/.style={
            draw, rounded corners=3pt, minimum height=0.8cm,
            minimum width=2.0cm, align=center, font=\small,
            fill=green!10, line width=0.8pt, draw=green!50!black
        },
        arrow/.style={-{Stealth[length=5pt]}, line width=0.6pt},
    ]

    % Top row
    \node[box] at (0, 0) (raw) {Raw\\[-1pt]Projections};
    \node[box] at (3.2, 0) (preproc) {Dark/Flood\\[-1pt]Correction};
    \node[box] at (6.4, 0) (log) {Log\\[-1pt]Transform};
    \node[box] at (9.6, 0) (cosine) {Cosine\\[-1pt]Weighting};

    \draw[arrow] (raw) -- (preproc);
    \draw[arrow] (preproc) -- (log);
    \draw[arrow] (log) -- (cosine);

    % Bottom row
    \node[box] at (1.6, -1.8) (parker) {Parker\\[-1pt]Weighting};
    \node[highlightbox] at (4.8, -1.8) (ramp) {Ramp\\[-1pt]Filter};
    \node[box] at (8.0, -1.8) (bp) {Back-\\[-1pt]projection};
    \node[box] at (11.2, -1.8) (hu) {HU\\[-1pt]Calibration};

    \draw[arrow] (cosine.south) -- ++(0, -0.5) -| (parker.north);
    \draw[arrow] (parker) -- (ramp);
    \draw[arrow] (ramp) -- (bp);
    \draw[arrow] (bp) -- (hu);

    \end{tikzpicture}
    \end{center}
    \caption{FDK reconstruction pipeline for short-scan cone-beam micro-CT. The ramp filter step (highlighted) is the only step varied in this study.}
    \label{fig:pipeline}
\end{figure}
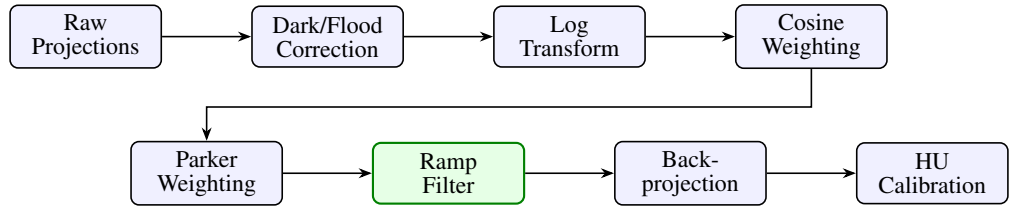

\noindent Figure~\ref{fig:pipeline} summarises this pipeline. Step~4 (ramp filtering) is the only step varied in this study; sixteen configurations were evaluated by combining four filter kernels (ramp, Shepp-Logan, cosine, Hamming) with four cutoff frequencies (1.0, 0.8, 0.6, and $0.379 \times$~Nyquist).

\subsection{Image Quality Metrics}\label{sec:metrics}

\subsubsection{Modulation Transfer Function}

The MTF was measured from the slanted-edge insert using an ISO~12233-style~\cite{iso_12233_2017} oversampled edge technique with $4\times$ subpixel alignment. The line spread function was obtained by differentiation of the edge response function and fitted with a Gaussian; the MTF was computed as the normalised magnitude of its Fourier transform. Spatial resolution is reported as the frequencies at which the MTF falls to 50\% and 10\% of its peak value (MTF$_{50}$ and MTF$_{10}$).

\subsubsection{Noise Power Spectrum}

The NPS was measured following ICRU Report~87~\cite{icru_ct_2012} from eight circular ROIs (radius = 66~pixels) arranged in a ring pattern (\ang{0} to \ang{315} in \ang{45} steps) over 16~axial slices in the homogeneous water region of the phantom. A third-order polynomial was subtracted from each ROI to remove low-frequency trends. The 2D NPS was computed as $|\mathrm{FFT}|^2$, normalised by $\Delta x^2 / (\pi r^2 N_\mathrm{ROI})$, and radially averaged. The integrated NPS was obtained by 2D Simpson integration.

\subsubsection{Detectability Index}

Task-based detectability was quantified using the non-prewhitening matched filter (NPW) observer model~\cite{aapm_tg233_2019}. The task function was a circular disc of contrast $\Delta C$ and radius $R$: $T(f) = \Delta C \, \pi R^2 \, 2J_1(2\pi R f) / (2\pi R f)$. The detectability index was computed as
\begin{equation}\label{eq:dprime}
    d'^2 = \frac{\left[\int_0^{f_\mathrm{max}} |T(f)|^2 \, \mathrm{MTF}^2(f) \, f \, df\right]^2}{\int_0^{f_\mathrm{max}} |T(f)|^2 \, \mathrm{MTF}^2(f) \, \mathrm{NPS}(f) \, f \, df}
\end{equation}
at two contrast levels: $\Delta C = \SI{500}{HU}$ (high-contrast tasks, e.g., iodine-enhanced vessels or bone--soft-tissue boundaries) and $\Delta C = \SI{100}{HU}$ (unenhanced soft-tissue lesion detection). Disc diameters spanned 0.1--\SI{3.5}{mm} at \SI{500}{HU} and 0.1--\SI{9}{mm} at \SI{100}{HU}, ensuring every configuration crosses the Rose threshold. The Rose criterion~\cite{rose_vision_1973} ($d' = 3$) was used to define the minimum detectable disc diameter for each configuration at each contrast level.

\newpage
%% ============================================================
\section{Results}\label{sec:results}
%% ============================================================

\subsection{MTF, NPS, and Detectability}\label{sec:quantitative}

\begin{figure}[H]
    \begin{center}
    \includegraphics[width=\textwidth]{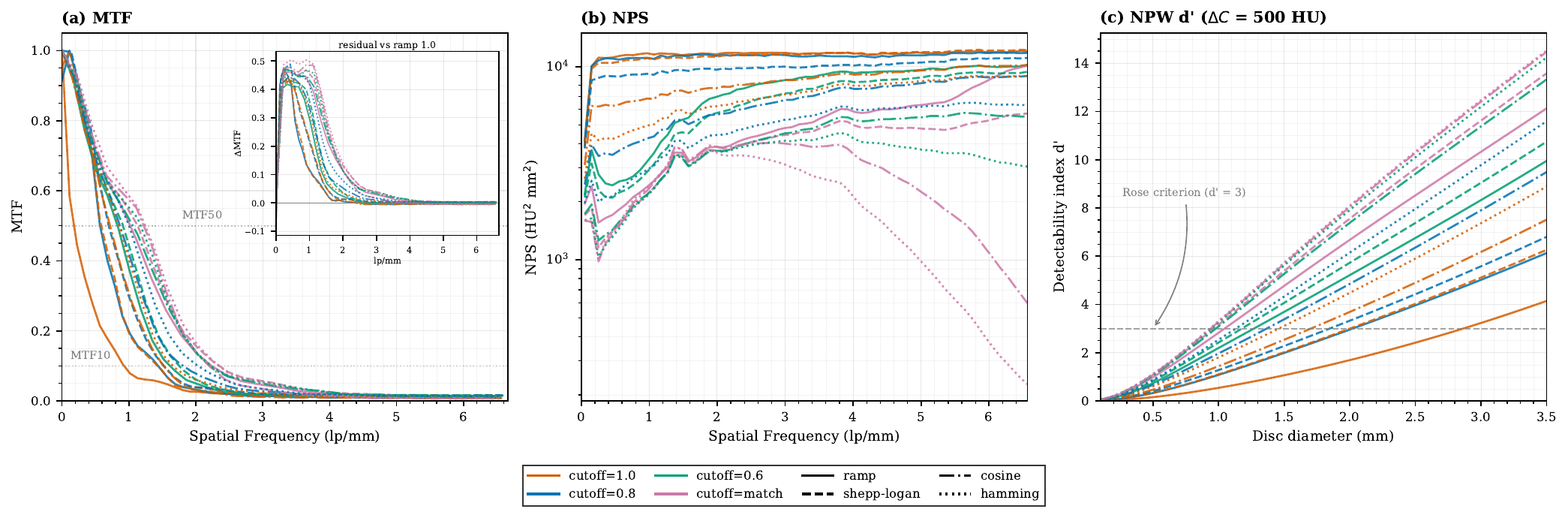}
    \end{center}
    \caption{Image quality metrics for all sixteen filter configurations on the image quality phantom (short-scan, \ang{193}). (a)~Modulation transfer function; inset shows residual relative to ramp at $f_c = 1.0$. (b)~Noise power spectrum. (c)~NPW detectability index at $\Delta C = \SI{500}{HU}$; dashed horizontal line marks the Rose criterion ($d' = 3$). Line styles denote filter kernel (solid: ramp, dashed: Shepp-Logan, dash-dot: cosine, dotted: Hamming); colours denote cutoff frequency.}
    \label{fig:metrics}
\end{figure}

Figure~\ref{fig:metrics} presents the MTF, NPS, and NPW~$d'$ for all sixteen filter configurations on the image quality phantom.

The MTF curves (panel~a) show that both the choice of window and the cutoff frequency affect the measured spatial resolution; the inset shows the residual MTF of each configuration relative to the ramp at $f_c = 1.0$. Table~\ref{tab:mtf10} lists the MTF$_{10}$ values for all sixteen configurations.

The NPS (panel~b) shows variation in both the magnitude and spectral shape of image noise across configurations. Configurations with higher cutoffs show elevated high-frequency noise power. Table~\ref{tab:nps} lists the integrated NPS for each configuration.

The NPW~$d'$ (panel~c) combines both effects into a single task-relevant metric. Table~\ref{tab:rose} lists the Rose criterion~\cite{rose_vision_1973} crossing diameters at $\Delta C = \SI{500}{HU}$; Table~\ref{tab:rose_lowc} lists the corresponding crossings at $\Delta C = \SI{100}{HU}$, a lower-contrast regime in which detectability requires substantially larger lesion diameters across all configurations.

\begin{table}[H]
\centering
\caption{MTF$_{10}$ (lp/mm) for each filter kernel and cutoff frequency. Best values in bold.}
\label{tab:mtf10}
\small
\begin{tabular}{lcccc}
\toprule
Cutoff & Ramp & Shepp-Logan & Cosine & Hamming \\
\midrule
1.0   & 0.93 & 1.43 & 1.52 & 1.66 \\
0.8   & 1.44 & 1.52 & 1.81 & 2.03 \\
0.6   & 1.62 & 1.75 & 2.21 & 2.23 \\
Match & 2.24 & \textbf{2.35} & 2.20 & \textbf{2.35} \\
\bottomrule
\end{tabular}
\end{table}

\begin{table}[H]
\centering
\caption{Integrated NPS (HU$^2$) for each filter kernel and cutoff frequency. Best value in bold.}
\label{tab:nps}
\small
\begin{tabular}{lcccc}
\toprule
Cutoff & Ramp & Shepp-Logan & Cosine & Hamming \\
\midrule
1.0   & 75\,670 & 74\,633 & 55\,174 & 46\,160 \\
0.8   & 73\,795 & 64\,970 & 43\,543 & 32\,781 \\
0.6   & 49\,783 & 44\,113 & 27\,895 & 22\,105 \\
Match & 33\,354 & 26\,458 & 17\,742 & \textbf{13\,259} \\
\bottomrule
\end{tabular}
\end{table}

\begin{table}[H]
\centering
\caption{Rose criterion crossing diameter (mm) at $\Delta C = \SI{500}{HU}$: the smallest disc diameter for which $d' \geq 3$. Smaller values indicate better detectability; best value in bold.}
\label{tab:rose}
\small
\begin{tabular}{lcccc}
\toprule
Cutoff & Ramp & Shepp-Logan & Cosine & Hamming \\
\midrule
1.0   & 2.86 & 2.00 & 1.70 & 1.46 \\
0.8   & 2.02 & 1.85 & 1.37 & 1.13 \\
0.6   & 1.27 & 1.18 & 0.98 & 0.94 \\
Match & 1.04 & 0.96 & 0.94 & \textbf{0.93} \\
\bottomrule
\end{tabular}
\end{table}

\begin{table}[H]
\centering
\caption{Rose criterion crossing diameter (mm) at $\Delta C = \SI{100}{HU}$: the low-contrast counterpart to Table~\ref{tab:rose}. Smaller values indicate better detectability; best value in bold.}
\label{tab:rose_lowc}
\small
\begin{tabular}{lcccc}
\toprule
Cutoff & Ramp & Shepp-Logan & Cosine & Hamming \\
\midrule
1.0   & 7.74 & 6.78 & 5.97 & 5.37 \\
0.8   & 6.96 & 6.43 & 5.13 & 4.41 \\
0.6   & 4.93 & 4.68 & 3.93 & 3.69 \\
Match & 4.28 & 3.86 & 3.63 & \textbf{3.62} \\
\bottomrule
\end{tabular}
\end{table}

\subsection{Spatial Resolution vs Cutoff}\label{sec:resolution_cutoff}

\begin{figure}[H]
    \begin{center}
    \includegraphics[width=\textwidth]{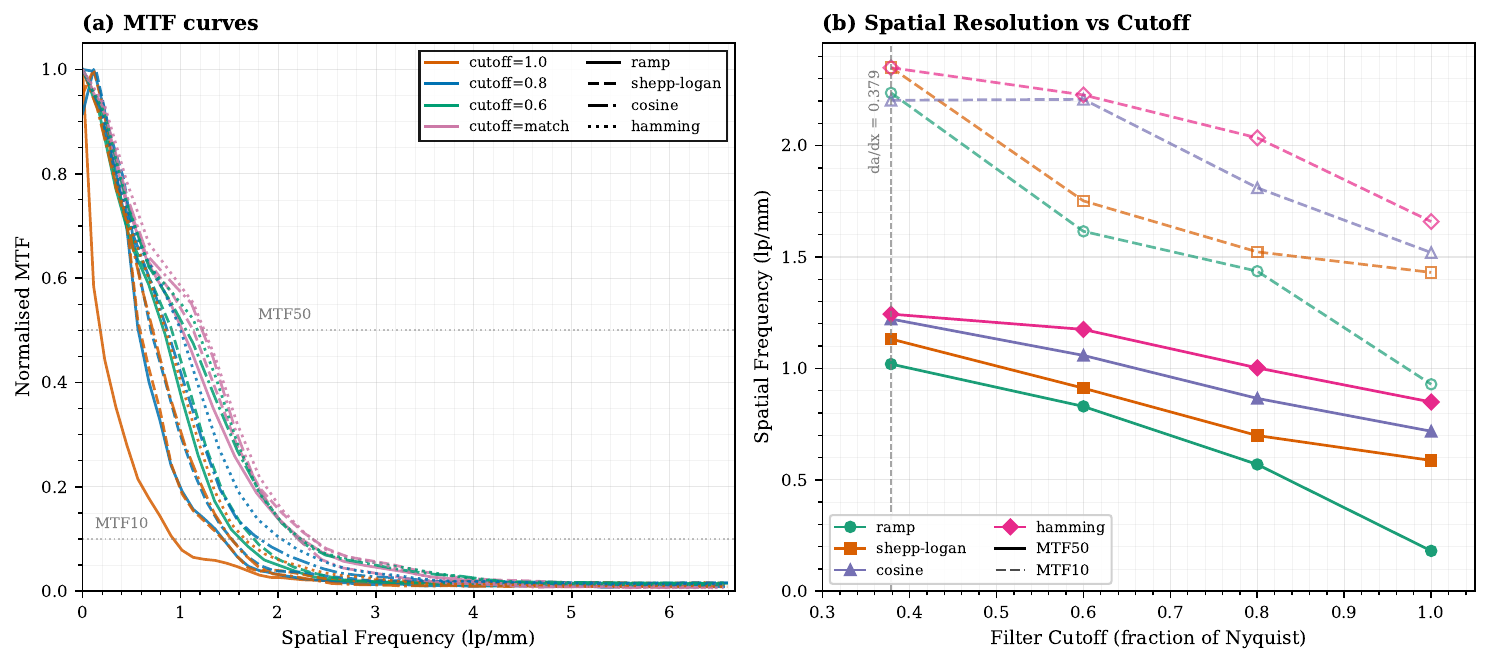}
    \end{center}
    \caption{Spatial resolution as a function of filter configuration. (a)~MTF curves for all sixteen configurations, with MTF$_{50}$ and MTF$_{10}$ thresholds indicated. (b)~MTF$_{50}$ (solid markers) and MTF$_{10}$ (open markers) vs cutoff frequency for each filter kernel. The vertical dashed line marks the matched cutoff $d_a/\Delta x = 0.379$.}
    \label{fig:mtf_cutoff}
\end{figure}

Figure~\ref{fig:mtf_cutoff} provides a complementary view of how spatial resolution varies with filter configuration. Panel~(a) shows the full MTF curves for all sixteen configurations. Panel~(b) plots MTF$_{50}$ and MTF$_{10}$ as a function of cutoff frequency for each filter kernel, showing the continuous relationship between cutoff and measured spatial resolution. The matched cutoff $d_a / \Delta x = 0.379$ is marked for reference.

\subsection{Qualitative Comparison on Mouse Lung}\label{sec:qualitative}

\begin{figure}[H]
    \begin{center}
    \includegraphics[width=0.75\textwidth]{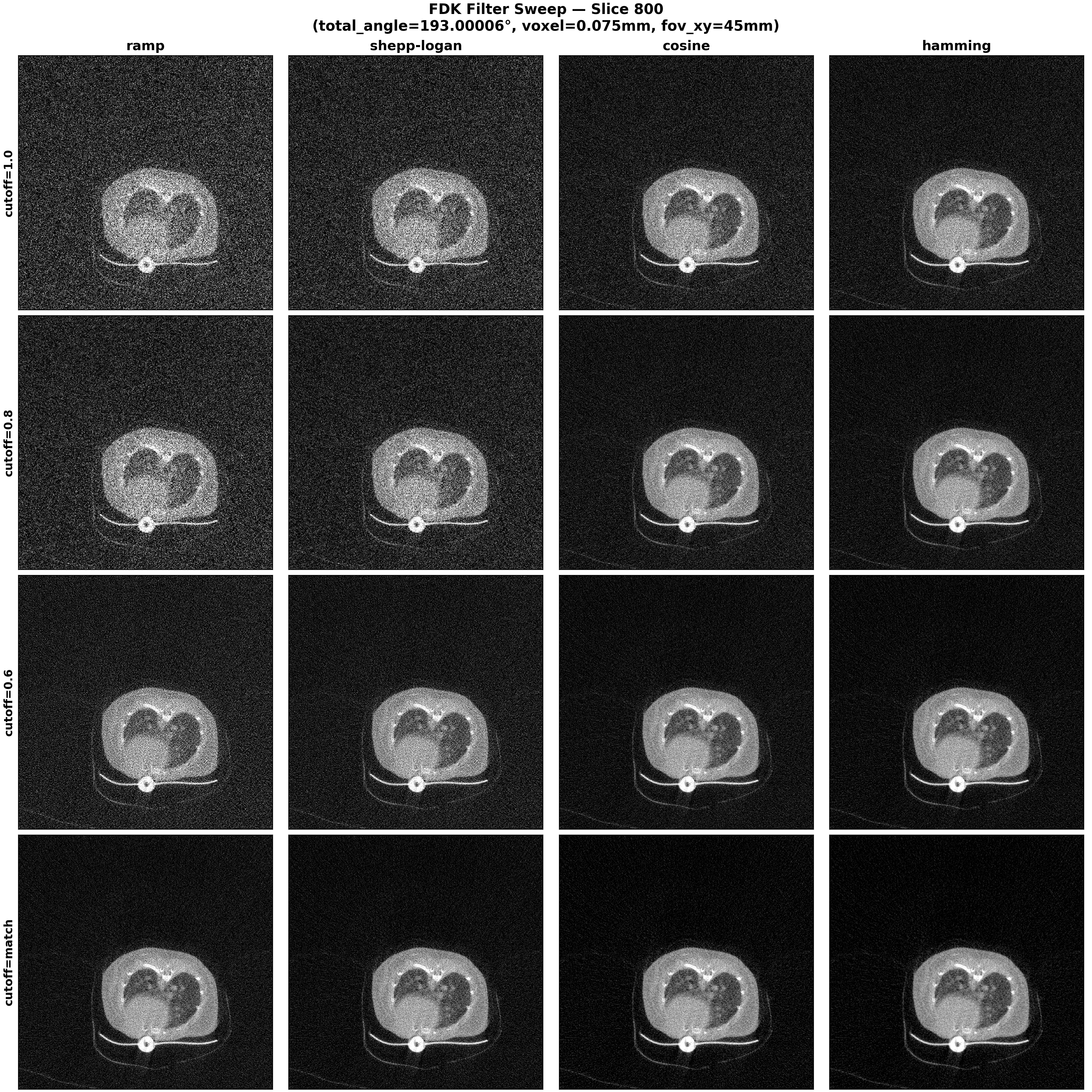}
    \end{center}
    \caption{Mouse lung reconstructions (axial slice) for all sixteen filter configurations, displayed with identical window/level settings. Columns: ramp, Shepp-Logan, cosine, Hamming. Rows: $f_c = 1.0$, 0.8, 0.6, matched ($0.379 \times$~Nyquist).}
    \label{fig:mouse}
\end{figure}

Figure~\ref{fig:mouse} presents axial slices of a mouse lung reconstructed with each of the sixteen configurations. The grid spans filter kernels (columns) and cutoff frequencies (rows), with all images displayed at identical window and level settings. The visual differences in noise texture and structural detail across configurations complement the quantitative metrics presented above.

%% ============================================================
\section{Discussion}\label{sec:discussion}
%% ============================================================

The data presented here show that filter kernel and cutoff frequency selection substantially affects all three image quality metrics. The NPS data in particular illustrate that the spectral distribution of noise---not just its magnitude---varies with filter configuration: configurations with higher cutoffs and weaker apodization concentrate more noise power at higher spatial frequencies, while stronger apodization reshapes the noise spectrum toward a flatter profile. The NPW~$d'$ captures how these spectral characteristics combine for a given detection task, providing a single metric that accounts for both signal transfer and noise.

The matched cutoff $f_c = d_a / \Delta x$ provides a physically motivated default by limiting the filter bandwidth to frequencies that are meaningfully sampled given the detector pixel pitch and reconstruction voxel size. This is the configuration used in our companion benchmarking study~\cite{wiegmann_fdk_benchmark_2026}, where the Hamming kernel at matched cutoff was adopted for the FDK pipeline.

These results are specific to the eXplore CT~120 geometry; the matched cutoff value will differ for systems with different detector pixel pitches or reconstruction voxel sizes. However, the general trends with window type and cutoff frequency are expected to hold for other scanners using the FDK algorithm. All quantitative metrics were measured on a single phantom; performance on other imaging tasks and specimens may differ.

%% ============================================================
\section{Conclusion}\label{sec:conclusion}
%% ============================================================

This note provides a systematic visual and quantitative reference for FDK filter kernel and cutoff frequency selection in preclinical cone-beam micro-CT. The sixteen configurations evaluated span a wide range of image quality characteristics, and the data are presented to assist groups in selecting filter parameters appropriate to their imaging task.

% References
\bibliographystyle{unsrtnat}
\bibliography{fdk_filter_sweep_paper}

\newpage

\section*{Acknowledgments}
This work was supported by the BC Lung Foundation.

\section*{Author Contributions}
Falk L. Wiegmann and Nancy L. Ford contributed to the research direction and conceptualisation. Falk L. Wiegmann developed the reconstruction pipeline, performed the analysis, and wrote the manuscript. Nancy L. Ford supervised the research, secured funding, provided critical review, and edited the manuscript.

\section*{Competing Interests}
The authors declare no competing interests.

\section*{Data Availability}
The reconstruction pipeline code is publicly available at \url{https://github.com/UBC-Ford-lab/eXplore_CT_120_fdk_reconstruction_algorithm}.
The scan data are available from the corresponding author on reasonable request.

\section*{Ethics Statement}
No animal experiments were conducted as part of this study. The image quality phantom data were acquired exclusively for evaluation purposes. The \textit{in vivo} mouse lung micro-CT data were collected as part of previously published studies~\cite{ford_respiratory_2025, ford_spie_2023} under protocols approved by the University of British Columbia Animal Care Committee (Protocol No.\ A21-0060, approved August 31, 2021) and performed in accordance with the ARRIVE guidelines~\cite{percie_du_sert_arrive_2020}. The imaging data were re-used here for reconstruction evaluation only; no animals were imaged, handled, or subjected to any procedures as part of this work.

\section*{Use of AI Tools}
Claude (Anthropic) was used to assist with manuscript preparation. All AI-generated content was reviewed, verified, and revised by the authors, who take full responsibility for the final manuscript.

\end{document}